# Morphological Type Dependence in the Tully-Fisher Relationship

David G. Russell

Owego Free Academy, Owego, NY 13827 USA

russeld1@oagw.stier.org

## Abstract

The Tully-Fisher relationship is subject to morphological type dependence such that galaxies of morphology similar to ScI galaxies and Seyfert galaxies are more luminous at a given rotational velocity than galaxies of other morphological classification. This effect is most prevalent in the B-band. It is shown that the type effect is not simply an artifact of the calibrator sample, but is also present in cluster samples. The type effect is corrected by creating type dependent Tully-Fisher relations for ScI group galaxies and Sb/ScIII group galaxies. It is shown that with single calibrations, the distances to ScI group galaxies are systematically underestimated while the distances to Sb/ScIII group galaxies are systematically overestimated. Tully-Fisher slope and scatter is also considered in the context of Type Dependent Tully-Fisher relations. It is concluded that the use of type dependent Tully-Fisher relations provide significant improvement in the distances to individual galaxies and refined distances to clusters of galaxies.

*Subject headings*: cosmology: distance scale – galaxies: fundamental parameters – distances

## Introduction

The Tully-Fisher relation (Tully&Fisher 1977) is the most important secondary distance indicator for determining distances to large samples of spiral galaxies. One contributor to scatter in the Tully-Fisher Relation (hereafter TFR) is morphological type dependence. Roberts (1978) found evidence that the slope of the TFR may depend upon morphological type. Rubin et al (1985) found that late type spiral galaxies are more luminous at a given rotational velocity than early type spiral galaxies in both the B-band and the H-band – a result that seems to be confirmed by additional studies (Giraud 1986; Giovanelli et al 1997; Corteau et al 2003). Sandage (1996,1999a,1999b) found evidence that both morphological T-type and luminosity class type are related to scatter in the TFR. Type dependence in the diameter TFR has also been found (Theureau et al 1997). Russell (2002,2003) found that the calibrator samples exhibit type dependence that is related to both morphological type and luminosity class. The cumulative results make the conclusion that there is morphological type dependence in the TFR not only in the blue bands, but also in near-infared bands undeniable (Corteau et al 2003).

Efforts to correct for type dependence have varied. Giovanelli et al (1997) made straight additive corrections to magnitudes for types early than Sb and for Sb galaxies. Sandage (1996, 1999a) corrected for both Hubble T-type and luminosity class with the relation: 0.23T + 0.5L. Another approach which has been advocated as a method to avoid the influence of morphological type dependence in calculation of individual galaxy distances is the method of "sosies" (Paturel 1984; Paturel et al 1998; Bottinelli et al 1999). In the sosie method galaxies are identified which have the same morphological type, inclination, and rotational velocity as a calibrator galaxy with a reliable zero point. The distance modulus to a galaxy may then be calculated by with the following relation:

$$\mu = \mu(calib) + Btc - Btc\ (calib) \quad \text{(Paturel et al 1998)} \quad (1)$$

Paturel et al argue that the method of sosies has an advantage in that corrections for inclination effects and morphological type dependence will have no effect on the final result because these factors are the same for the calibrator and the sosie galaxies. The largest disadvantage to the sosie method is the error in the calibrator distance because the sosie method guarantees that a typical error of 1σ in the cepheid distance will be systematically carried through the entire sosie sample.

Russell (2002,2003) showed that both morphological type and luminosity class can be factored into TFR equations and adopted a method Type Dependent Tully-Fisher Relations (hereafter TD-TFR).   Figure 1 is a plot of the TFR in the B-band for the calibrator sample used by Russell (2003).  Error bars are rms uncertainty in the logarithm of the rotational velocity and uncorrected B-band magnitude for the calibrator sample as provided in the LEDA database. The filled circles are galaxies of morphological types SbcI,SbcI-II,ScI, ScI-II, and Seyfert galaxies (hereafter ScI group).   It is clearly seen that the galaxies of the ScI group are more luminous at a given rotational velocity than the Sab/Sb/Sbc/Sc galaxies of luminosity classes that do not place them in the ScI group. The Sab/Sb/Sbc/Sc II-III to III-IV galaxies will hereafter be referred to as the Sb/ScIII group.

Russell (2003) adopted two calibrator samples in which ScI group galaxies provided one calibrator sample of 12 galaxies and Sb/ScIII group galaxies provided a second calibrator sample of 15 galaxies.   The resulting B-band type dependent equations are (see Russell 2003 for calibrator data):

m-M   =   19.85(+/- 0.16) + 5.24(+/- 0.10)(log Vmax - 2.2) + Btc   Sb/ScIII group   (2)

m-M   =   20.44(+/-0.09) + 4.91(+/-0.20) (log Vmax - 2.2) + Btc    ScI group        (3)

It should be noted that all galaxies utilized in this analysis are Hubble T-type 6 and earlier and with rotational velocities greater than 95 km s$^{-1}$.  It would therefore be inappropriate to apply equations 2 and 3 to later type or slower rotating spirals.  In equations 2 and 3, Btc is the corrected B-band magnitude from the LEDA database and log Vmax is the logarithm of the rotational velocity of the galaxy.

Note that the TD-TFR equations have very similar slopes such that difference in distance moduli primarily results from a zero point offset that is 0.59 mag. Due to the systematic nature of the type effect, ScI galaxies will have underestimated TFR distances and Sb/ScIII group galaxies will have overestimated TFR distances if the type effect is not accounted for (Russell 2003). The purpose of this paper is to demonstrate that the TD-TFR's are not merely a product of the calibrator sample but that the same effects are readily evident in clusters of galaxies.

## 2. Cluster samples

The sample of clusters listed in Table 1 was selected from the cluster sample of Giovanelli et al (1997a – hereafter G97) and supplemented with two clusters from the survey of Mathewson & Ford (1996 – hereafter MF96) and one group from the LEDA database. Individual galaxies selected from each cluster were restricted to morphological type Scd and earlier and to the surface brightness range (bri25 in LEDA) of the calibrator sample. Rotational velocities were taken from G97 for A2197/99, A262, NGC 338/507, Pegasus, A2634, Coma, Antlia, and Hydra. Rotational velocities for the Ursa Major cluster were Vflat from Verheijen (2001). For the ESO 234-22 and 553-26 groups rotational velocities were taken from MF96. Finally, rotational velocities for the NGC 1030 group and all corrected B-band magnitudes were taken from the LEDA database.

It should be noted that there is reason for caution when using a heterogeneous database such as LEDA and when utilizing rotational velocities from multiple sources. In figure 2, rotational velocities from the LEDA database and MF96 are compared with rotational velocities from G97 for galaxies in the Antlia and Hydra clusters. It can be seen that all three sets of rotational velocities are consistent. The 1σ scatter in the logarithm of the rotational velocity between G97 and MF96 is 0.022 which produces distance modulus differences of ~0.108 to 0.115 mag with the slopes in equations 2 and 3. The 1σ scatter between G97 and LEDA is 0.024 which produces distance modulus differences of ~0.118 to 0.126 mag.

*It must also be noted that the LEDA team does not homogenize corrected magnitudes and rotational velocities published from other sources into their database.* Raw data is collected for the LEDA database, homogenized, and transformed to the same standard. All magnitudes and rotational velocities are then corrected for inclination, absorption, and K-effect with a homogeneous system of corrections (Theureau 2004, private communication).

## 3. Visibility of the Type effect in the cluster samples

### 3.1 Results for the cluster sample

It was noted in Russell (2003) that if the type effect is not accounted for ScI group galaxies will have systematically underestimated Tully-Fisher distances while Sb/ScIII group galaxies will have systematically overestimated distances. This was illustrated with the Ursa Major cluster for which it was shown that when the type dependent equations were used mean distance moduli of ScI group and Sb/ScIII group galaxies agreed within 0.06 mag (B-band). However, with a single calibration the Sb/ScIII group mean distance modulus was larger than the ScI group by 0.45 mag.

A single slope was fit to all 27 calibrators and the single TFR derived for the B-band is:

$$m\text{-}M \quad = \quad 20.06(+/\text{-} 0.30) + 5.75(\log V_{max} - 2.2) + B_{tc} \quad \text{(B-band)} \qquad (4)$$

For each cluster in Table 1, distance moduli were calculated using the TD-TFR (equations 2 and 3) as well as using the single TFR (equation 4). Column 1 is the cluster; column 2 is the number of galaxies in the sample; column 3 is the mean cluster distance modulus with the TD-TFR; column 4 is the 1 scatter of the individual galaxy distance moduli around the mean cluster distance; column 5 is the mean cluster distance modulus using the traditional single TFR; column 6 is the 1 scatter of the individual galaxy distance moduli around the mean cluster distance. In figure 3, the B-band distance moduli of individual ScI group galaxies from the cluster samples are plotted against mean distance moduli for the Sb/ScIII group galaxies within the respective clusters. The open circles are the ScI distances with the TD-TFR while the filled circles are the ScI distances with the traditional single TFR. The solid line in figure 3 represents equal distance moduli for both the Sb/ScIII and ScI groups. The error bars are the uncertainty in distance moduli for the cluster samples derived from the errors in B-band magnitudes and rotational velocities. It is clearly seen that the individual ScI galaxies cluster around the mean distance of the Sb/ScIII group galaxies when the TD-TFR is used. *However, when a single calibration is used every one of the ScI group galaxies has a Tully-Fisher distance modulus that is less than the mean distance modulus of the Sb/ScIII galaxies in the cluster.*

The triangles in figure 3 are for five close pairs of galaxies with one ScI group and one Sb/ScIII group galaxy (Table 2). As with the clusters, the two galaxies in the pairs have very close agreement of distance moduli with the TD-TFR, but the ScI galaxies are systematically less distant than the Sb/ScIII galaxies with the single calibration.

Figures 3 clearly establishes that there is a significant type effect in the B-band TFR which is not limited to the calibrator sample. If this type effect is not accounted for ScI group galaxies will have systematically underestimated TFR distances while Sb/ScIII group galaxies will have systematically overestimated TFR distances. The ScI group distance moduli in this sample are systematically smaller than the Sb/ScIII group distance moduli by 0.46 mag with the traditional single TFR. Note that the uncertainty from measurement errors (error bars on figure 3) is only 0.27 mag.

## 3.2 Comments on individual clusters and galaxies

*Antlia & Hydra* - The ScI group calibrator sample includes the Sb type Seyfert galaxies NGC 1365, NGC 3627, and NGC 4258. Antlia contains the SBbII Seyfert galaxy IC 2560 which has a TD-TFR distance modulus of 32.49 (B-band) - identical to the mean distance modulus for the cluster. If the single calibration was used the distance modulus of IC 2560 would be the smallest in the cluster at 32.19 and 0.37 mag less than the mean of 32.56 given from the single calibration. Thus IC 2560 supports the result from the calibrator sample that Sb type Seyfert's belong in the ScI group.

Antlia and Hydra also have as members the galaxies NGC 3095 (SBcII), ESO 501-75 (ScII), and NGC 3464 (SBcII). Each of these galaxies is a faster rotator that has a distance modulus close to the cluster mean if included in the ScI group rather than the Sb/ScIII group. This is a strong indication that faster rotating ScII galaxies should be included in the ScI group TD-TFR.

*A2197/99* – NGC 6195 is an SbI galaxy in the A2197/99 cluster that has a distance modulus only 0.12 mag less than the cluster mean if it is assigned to the ScI group. In the Sb/ScIII group NGC 6195 would have a distance modulus of 34.55 which is 0.63 mag less than the cluster mean. This suggests that Sb galaxies of luminosity class I may also be members of the ScI group TD-TFR.

*A2634* - This was the only cluster in the sample with no galaxies that are members of the ScI group. The effect of type dependence is particularly important for a cluster such as A2634 with no ScI group galaxies. With the TD-TFR A2634 has a distance modulus of 34.82. Using the single calibration the distance modulus is increased to 35.07. The difference amounts to about a 10% change in the value of the Hubble Constant indicated by this cluster. While A2634 is the most extreme case, in every cluster in our sample, the use of a single calibration leads to an increase in cluster distance and therefore a lower value for the Hubble Constant.

## 3.3 Classification errors

The biggest concern with the use of TD-TFR is the possibility that some galaxies will be incorrectly classified and therefore would fall into the wrong group. Fortunately, the narrow arms of SbcI and ScI galaxies are distinctive and thus it is unlikely that galaxies would be incorrectly assigned an ScI group morphology. However, it is possible that some galaxies that should be included in the ScI group will be included in the Sb/ScIII group. This concern is greater at higher inclinations where the arm structure becomes more difficult to assign. For example, the calibrator NGC 1425 is classified as an SbII galaxy in the LEDA database, but has a zero point consistent with the ScI group and therefore was assigned to the ScI group for calibration of the TD-TFR. It should be noted that NGC 1425 is classified as an ScI-II galaxy in the RSA (Sandage&Tammann 1981) which suggests that the ScI group assignment is warranted.

Of the 152 galaxies in our sample, there were eight (5.3%) that had Sb/ScIII group morphology as classified in LEDA, but had TF distance moduli significantly smaller than the cluster mean using the Sb/ScIII TD-TFR. These galaxies are listed in Table 3 with the galaxies NGC 3095, ESO 501-75, NGC 3464, and NGC 6195 that were discussed above. For each of the eight galaxies discussed below the ScI group TD-TFR was used to calculate their distance and the mean distance of their cluster. It is encouraging that this rate of misclassification is consistent with the 27 TD-TFR calibrators for which one galaxy (NGC 1425 – 3.7%) was incorrectly classified in LEDA.

*NGC 3223* – This galaxy is classified as an SbI-II in LEDA so its luminosity class is consistent with the ScI group. The Hubble T-type in LEDA is 3.3 which is very close to the 3.5 that would be needed to make it an Sbc I-II galaxy. So within the errors of Hubble T-type classification, NGC 3223 appears to be a true member of the ScI group.

*NGC 3347* - This galaxy is classified as an SbI-II in LEDA with a Hubble T-type of 3.2. As with NGC 3223 its TF distance modulus is a better fit to the cluster mean if it is assigned to the ScI group.

*NGC 801 and NGC 818* - Both of these galaxies are Sc galaxies with no luminosity class assigned in LEDA. They are better fit to the mean cluster distance of A262 if they are assigned to the ScI group. Both galaxies have large rotational velocities - which are typical of ScI galaxies rather than ScIII galaxies. It is worth noting that NGC 801 has an inclination of 86 degrees which makes luminosity class determination extremely difficult.

*NGC 295, NGC 536, and PGC 10104* – All three of these galaxies are fast rotators classified as SBb galaxies in the LEDA database and better fit their cluster distances when assigned to the ScI group. It is noteworthy that five of the six Sb or Sbc galaxies that are mis-classified in LEDA are barred spirals.

*NGC 3318* – This galaxy is classified as an SBbcII-III and therefore is an unexpected member of the ScI group. What is distinctive about this galaxy is that it has a high surface brightness (bri25 in LEDA) compared with most of the galaxies in our sample. If assigned to the ScI group, NGC 3318 has a distance modulus of 32.52 (B-band) which is an excellent fit to the cluster mean.

Finally, it is worth noting that there is a common feature among the eight apparently incorrectly classified galaxies in Table 2. Each has a rotational velocity exceeding 197 km s$^{-1}$. It thus appears that faster rotating galaxies of Hubble T-types 3.1 to 5.0 are the most likely candidates for galaxies that could be incorrectly classified. However, such misclassifications make up only 5.3 % of our sample. They are easily identified in clusters from TF distance moduli that are noticeably smaller than the cluster mean and can therefore be reassigned to the correct TD-TFR group. It would be expected that in surveys that extend to greater distances, classification errors will become a greater concern when using Type Dependent Tully-Fisher relations. Classification errors of the types discussed in this section are more problematic in isolated galaxies because the distance cannot be checked against a group distance.

## 4.  Slope and Scatter

### 4.1  Scatter

Major Tully-Fisher studies have used cluster template relations to determine the slope of the TFR (Giovanelli et al 1997a,b; Tully & Pierce 2000; Sakai et al 2000).   In this study we have derived the Type Dependent Tully-Fisher Relations from the calibrator sample alone.  The calibrator sample is split into two smaller samples of 12 galaxies (ScI group) and 15 galaxies (Sb/ScIII group).   It must be noted that Tully&Pierce (2000) found that the slope and zero point derived from their cluster template relations was virtually identical to the slope and zero point derived from the calibrator sample alone. Thus it would appear that the smaller calibrator sample can be used to derive useful Tully-Fisher relations.

The $1\sigma$ zero point scatter of the TD-TFR is +/-0.09 mag  for the ScI group and  +/- 0.16 mag for the Sb/ScIII group. Note that this is the empirical scatter of the distance moduli relative to the calibrator distance moduli and does not account for measurement uncertainty (see error bars on figure 1).  The weighted mean scatter derived from $1\sigma$ scatter around mean cluster distance moduli is +/- 0.22 mag. for the cluster sample listed in Table 1.  The scatter of both the calibrators and the clusters is significantly smaller than the typical B-band scatter reported for studies that have used single TFR calibrations.  For example, Sakai et al reported a $1\sigma$ scatter of 0.45 mag in the B-band. Tully&Pierce (2000) report B-band scatter of 0.30 from the calibrators and 0.38 mag. from cluster templates.

While an effect from small numbers of calibrators on scatter cannot be avoided, the small scatter in this analysis is not simply a result of the smaller number of calibrators in each group, but is significantly improved from grouping galaxies according to morphological type.   To illustrate this point the entire calibrator sample was split into two samples of 13 and 14 galaxies – each sample containing 6 ScI group galaxies and the remainder galaxies from the Sb/ScIII group.   A least squares slope was fit to both samples in the B-band resulting in slopes of 6.05 and 5.60 from which new mean zero points were calculated.  The $1\sigma$ scatter derived from these two groups was 0.23 mag and 0.35 mag respectively.   When the entire calibrator sample was utilized for deriving a single TFR, the $1\sigma$ scatter was found to be 0.30 mag. which is the same as the B-band scatter reported by Tully&Pierce (2000) from their calibrator sample.  The fact that the scatter found by Tully&Pierce (2000) is recovered when a single calibration is applied to the 27 calibrators used in this study reinforces that the small scatter for the type dependent equations does not entirely result from small numbers.    The increase in TFR scatter resulting from single calibrations is also seen in Table 1 for the cluster samples which have a weighted mean scatter of +/- 0.28 mag with the single calibration and only +/- 0.22 mag with the TD-TFR.

4.2  Slope

   For this analysis, the TD-TFR slopes were derived directly from the smaller type dependent calibrator samples.  An alternative possibility would be to derive the slope from the complete calibrator sample and then derive separate type specific zero points for the individual groups.  The 27 calibrators give a slope of 5.75.  Using this slope the zero points would be 20.34 for the ScI group and 19.84 for the Sb/ScIII group – very close to the zero points found with the adopted slopes.

This approach to defining the slope is problematic because empirically, the calibrators indicate that ScI group galaxies are more luminous at a given rotational velocity than Sb/ScIII group galaxies. However, the slowest rotator among the ScI group calibrators is NGC 3198 for which log Vrot – 2.184. Among the Sb/ScIII group calibrators there are five galaxies with log Vrot from 1.989 to 2.146. If ScI group galaxies remain more luminous at a given rotational velocity for the slower rotators, then the absence of slower rotators among the ScI group calibrators will result in a steeper slope when the entire calibrator sample is utilized. In fact it is seen that the slope of equation 4 is steeper than the slopes of the TD-TFR.

Two tests were conducted to further illustrate this point. First, three slow rotating ScI group galaxies from Russell (2003) were added to the full sample of 27 calibrators. These galaxies were NGC 7610 (log v= 2.172), ESO 217-12 (log v= 2.134), and ESO 404-31 (log v=2.076). For the purposes of this test it is assumed that the TF distances determined in Russell (2003) may serve as calibrator distances. When these three galaxies are added to the full calibrator sample, the resulting sample of 30 galaxies has a slope of 5.36 as opposed to the 5.75 derived without the slow rotating ScI group galaxies.

As a second test, the calibrator sample was restricted to all galaxies with rotational velocities greater than log v=2.140 so that the Sb/ScIII group and ScI group galaxies covered the same rotational velocity range. This reduces the calibrator sample to 23 galaxies. The slope derived from this set is 5.35 – virtually identical to the slope derived when slower rotating ScI group galaxies were added. It therefore appears that the best single slope for the B-band TFR may be 5.35. Adopting this slope the TD-TFR zero points would be 20.40 ($1\sigma = 0.11$) for the ScI group and 19.85 ($1\sigma = 0.16$) for the Sb/ScIII group. It is important to note that this slope and zero point combination provides distance moduli within 0.10 mag of equations 2 and 3 for the full range of rotational velocities encompassed by the calibrator samples.

These tests indicate that slopes derived from cluster template relations may suffer from incomplete representation of both ScI and Sb/ScIII group galaxies across the rotational velocity range of the cluster sample. Slopes were derived from the cluster samples for A2634, A2197/99, A262, NGC 383/507, and Antlia. Since A2634 lacks ScI group galaxies it would be predicted that it should produce a shallower slope than the clusters with ScI group galaxies. A2634 sample indicates a slope of 5.14 compared with slopes of 9.76(A262), 5.70(NGC 383/507), and 6.26(Antlia) for clusters that have fast rotating ScI group galaxies. As found with the calibrator sample, the lack of slower rotating ScI group galaxies increases the slope of the Tully-Fisher relation. A2197/99 has two ScI group galaxies in the sample of 10 galaxies. These two galaxies are the fastest rotator (NGC 6195) and slowest rotator (cgcg 224-11) in the cluster. The slope for this cluster is 5.16. Here we see again that when the ScI group galaxies are found at both ends of the rotational velocity range the slope produced is consistent with the best slope derived from the calibrators.

As a final test the slopes of Antlia and NGC 383/507 were recalculated after removing the ScI group galaxies. The resulting slopes were shallower at 5.03 and 5.01 respectively. We conclude that determining the Tully-Fisher slope from cluster samples produces Tully-Fisher slopes too large if the ScI group galaxies do not cover the full rotational velocity range of the sample.

The slopes for each of the tests are listed in Table 4 and grouped as either "ScI unbiased" or "ScI biased".  It must be stressed that this grouping does not address the traditional cluster incompleteness bias (Teerikorpi 1987).  The classification of samples as ScI biased or unbiased is only for the purpose of identifying samples for which the ScI galaxies do not cover the full rotational velocity range of the cluster sample.  Also included in the ScI unbiased groups are those samples which contain only Sb/ScIII group galaxies.  It is noteworthy that  the "ScI unbiased" cluster samples indicate a remarkably small range in slope of 5.01 to 5.16.  In contrast the cluster samples with fast rotating ScI galaxies (ScI biased) indicated steeper slopes of 5.70 to 9.76.

## 5.  Conclusion

The results of this analysis indicate that the B-band Tully-Fisher relation can be significantly improved by utilizing Type Dependent Tully-Fisher relations which account for the systematic trend whereby galaxies of morphology similar to ScI galaxies are more luminous at a given rotational velocity than Sb/ScIII galaxies.   The use of TD-TFR was found to improve the TFR in the following ways:

1.  If the TD-TFR is not utilized, then ScI group galaxies have systematically underestimated TF distances while Sb/ScIII group galaxies have systematically overestimated TF distances.

2.  A single calibration will overestimate the slope of the TFR in both the calibrator sample and cluster templates if the cluster lacks ScI group galaxies over its full rotational velocity range.

3.  TF scatter is significantly reduced when the TD-TFR is used.  With single calibrations the $1\sigma$ scatter of the calibrators is 0.30 mag while the scatter is reduced to only 0.09 to 0.16 mag with the TD-TFR. In addition, scatter around mean cluster distance in galaxy clusters was found to be 0.06 mag smaller with the TD-TFR than when a single calibration is used.

Since galaxies may be grouped into the TD-TFR along clearly defined morphological lines with a low rate of morphological classification error, the results of this analysis suggest that the TD-TFR is a significant improvement over the standard single TFR in the B-band. Smaller but detectable gains in TFR accuracy may be expected if the TD-TFR is applied to R, I, and K' band TF relations.

Acknowledgements:

This research has made use of the Lyon-Meudon extragalactic Database (LEDA) compiled by the LEDA team at the CRAL-Observatoire de Lyon (France). I would also like to thank the anonymous referee for helpful suggestions.

Tables 1:  Cluster Sample Distance Moduli

| 1 | 2 | 3 | 4 | 5 | 6 |
|---|---|---|---|---|---|
| Cluster | # galaxies | m-M (B) | | m-M | |
| | | TD-TFR | | TFR | |
| Ursa Major | 18 | 31.36 | 0.38 | 31.38 | 0.44 |
| A2197/99 | 10 | 35.18 | 0.11 | 35.35 | 0.30 |
| A262 | 14 | 33.71 | 0.23 | 33.82 | 0.30 |
| N338/507 | 16 | 33.66 | 0.17 | 33.84 | 0.24 |
| Pegasus | 8 | 33.52 | 0.26 | 33.63 | 0.33 |
| A2634 | 16 | 34.82 | 0.15 | 35.07 | 0.15 |
| Coma | 19 | 34.58 | 0.30 | 34.76 | 0.33 |
| Antlia | 17 | 32.49 | 0.20 | 32.56 | 0.29 |
| Hydra | 9 | 33.20 | 0.13 | 33.28 | 0.23 |
| 234-22 | 9 | 33.83 | 0.16 | 33.86 | 0.24 |
| 553-26 | 11 | 34.46 | 0.16 | 34.63 | 0.14 |
| N1030 | 5 | 35.00 | 0.23 | 35.21 | 0.38 |

Table 2: Pairs with an ScI group and an Sb/ScIII group galaxy

| Galaxy | TF Group | TD-TFR(B) | Single TFR(B) |
|--------|----------|-----------|---------------|
| NGC 7537 | Sb/ScIII | 32.39 | 32.59 |
| NGC 7541 | ScI | 32.35 | 32.07 |
| | | | |
| ESO 545-13 | ScI | 34.69 | 34.44 |
| ESO 545-18 | Sb/ScIII | 34.81 | 35.01 |
| | | | |
| ESO 147-5 | ScI | 34.65 | 34.34 |
| ESO 147-10 | Sb/ScIII | 34.75 | 35.02 |
| | | | |
| ESO 52-20 | ScI | 34.00 | 33.63 |
| ESO 53-2 | Sb/ScIII | 34.01 | 34.21 |
| | | | |
| ESO 31-15 | Sb/ScIII | 34.69 | 34.94 |
| ESO 31-18 | ScI | 34.64 | 34.33 |

Table 3: ScI group galaxies with Sb/ScIII group morphology in LEDA

| Galaxy | Type | log Vrot | i | t | lc |
|--------|------|----------|----|-----|-----|
| NGC 3095 | SBcII | 2.324 | 59 | 5.1 | 3 |
| 501-75 | ScII | 2.256 | 61 | 5.1 | 3 |
| N3464 | SBcII | 2.290 | 51 | 5 | 2.9 |
| N6195 | SbI | 2.448 | 49 | 3 | 1 |
| N801 | Sc | 2.343 | 86 | 5 | |
| N818 | Sc | 2.366 | 71 | 4.8 | |
| N295 | SBb | 2.373 | 71 | 2.9 | |
| N536 | SBbII | 2.405 | 75 | 3.1 | 3 |
| N3223 | SbI-II | 2.429 | 51 | 3.3 | 2 |
| N3318 | SBbcII-III | 2.300 | 59 | 3.6 | 3.9 |
| N3347 | SBbI-II | 2.327 | 65 | 3.2 | 1.9 |
| PGC 10104 | SBb | 2.296 | 60 | 3 | |

Table 4:  B-band Tully Fisher slope

| Sample | Slope |
|--------|-------|
| ScI unbiased slopes: | |
| ScI group calibrators | 4.91 |
| Sb/ScIII group calibrators | 5.24 |
| All calibrators w/N7610,ESO 217-12, ESO 404-31 | 5.36 |
| Calibrators restricted to log Vrot >2.140 | 5.35 |
| A2634  (no ScI group galaxies in sample) | 5.14 |
| A2197/99 (ScI group galaxies cover cluster log Vrot range) | 5.16 |
| Antlia without ScI group galaxies | 5.03 |
| N383/507 without ScI group galaxies | 5.01 |
| | |
| ScI biased slopes: | |
| All 27 calibrators | 5.75 |
| A262   (includes ScI group galaxies) | 9.76 |
| N383/507  (includes ScI group galaxies) | 5.70 |
| Antlia  (includes ScI group galaxies) | 6.26 |

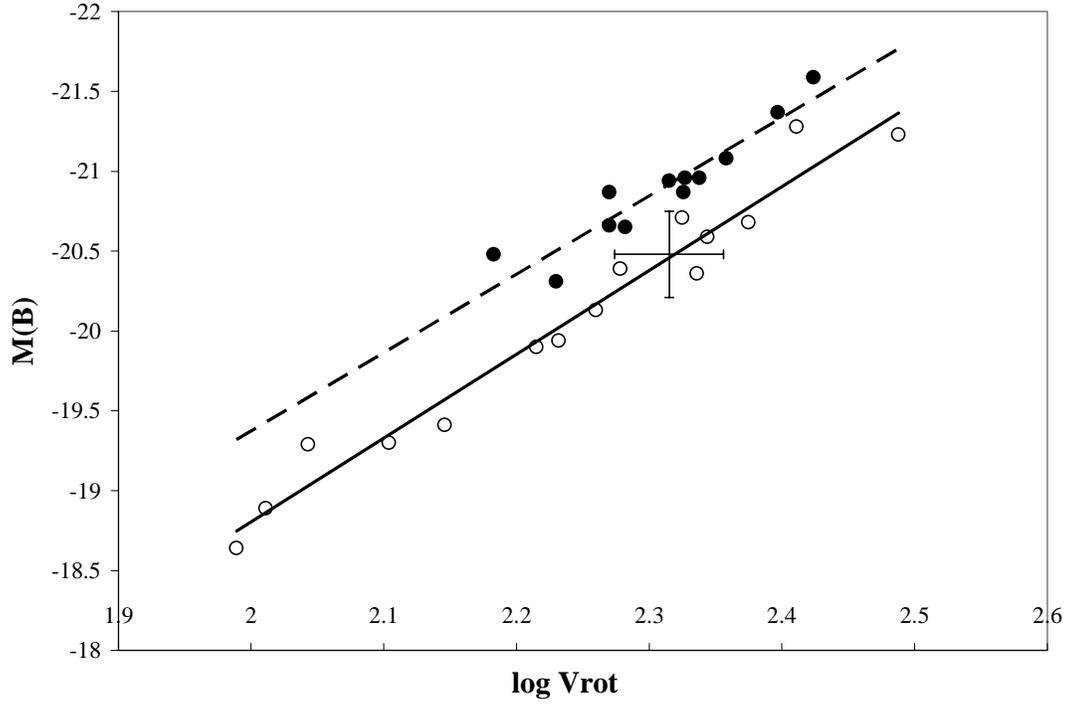

Figure 1 – B-band Calibration of Type Dependent Tully-Fisher Relation. Filled Circles are ScI group galaxies and dashed line is a least squares fit which gives a slope of 4.91. Open circles are Sb/ScIII galaxies and solid line is a least squares fit which gives a slope of 5.24. Error bars are $1\sigma$ uncertainty in magnitudes (+/- 0.27) and log Vrot (+/- 0.041) .

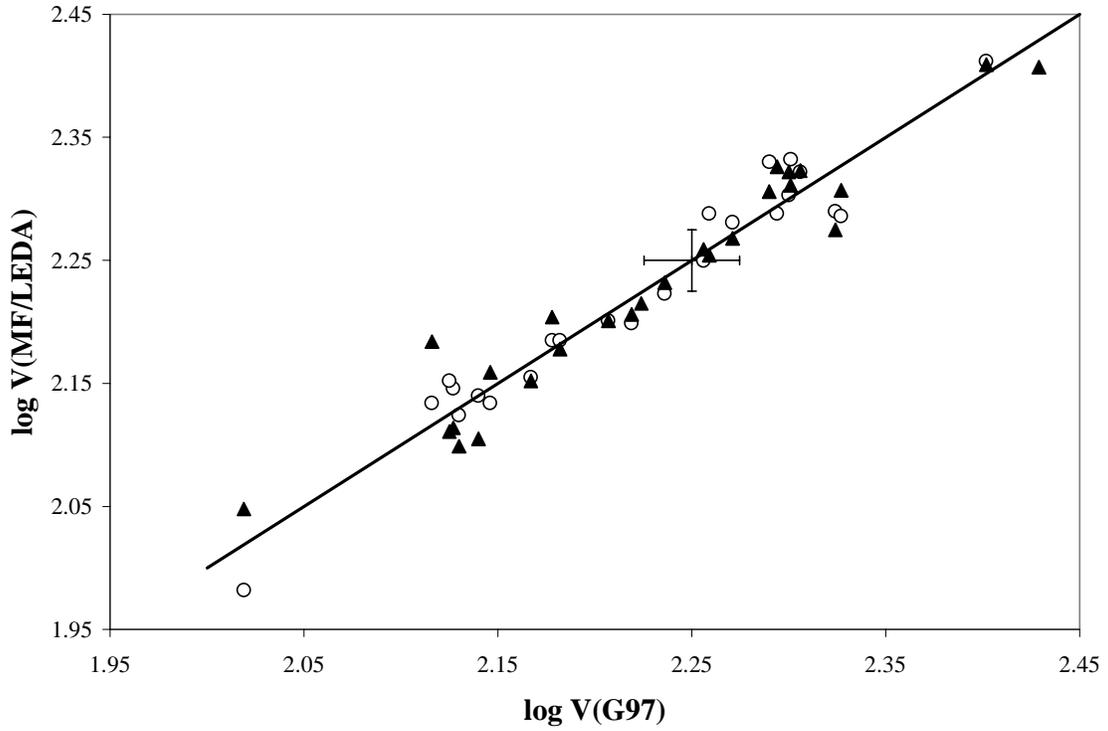

Figure 2 – Comparison of MF96 and LEDA rotational velocities with G97 rotational velocities. Open circles are MF96 rotational velocities and filled triangles are LEDA rotational velocities. Solid line represents LEDA and MF96 rotational velocities equal to the G97 velocities.

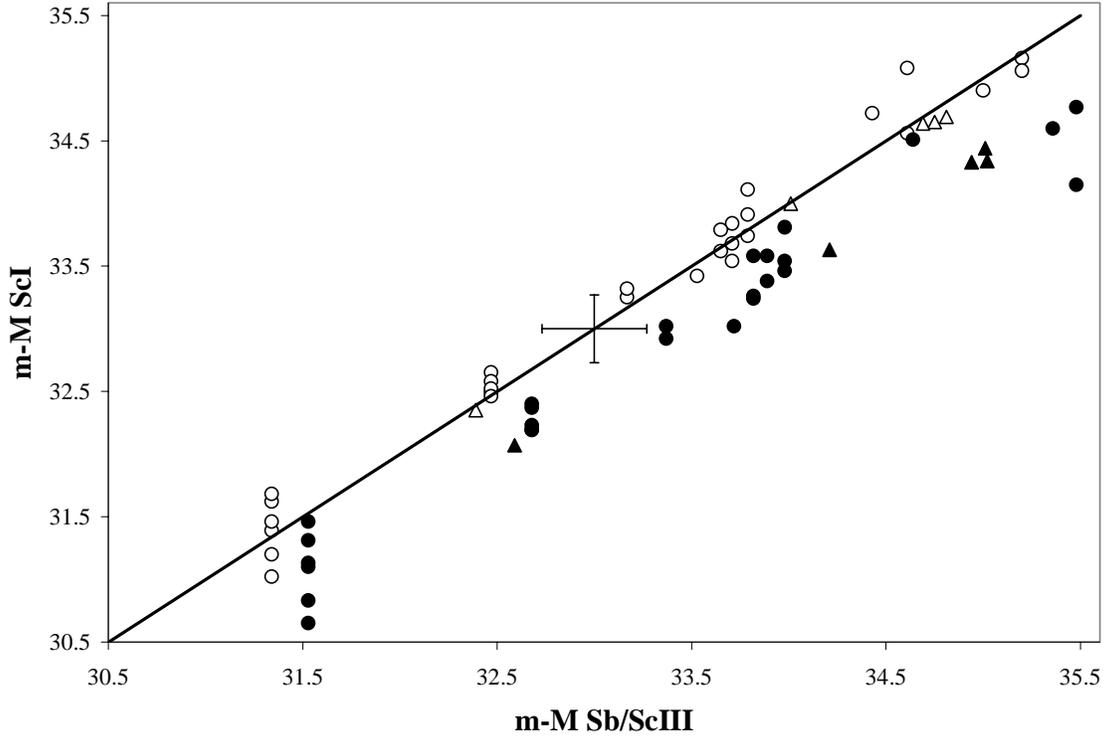

Figure 3 – Individual ScI group B-band distance moduli are compared with mean Sb/ScIII group distance moduli for the respective clusters. Open circles are ScI group distances using the Type Dependent Tully-Fisher relationship. Filled circles are the ScI group distances using the traditional single Tully-Fisher calibration. Solid line represents equal distances for ScI group and Sb/ScIII group galaxies. ScI distances are an excellent fit to the Sb/ScIII group mean distances with the TD-TFR but are systematically too small with the single calibration. Open Triangles are distances to ScI group galaxies in 5 pairs listed in Table 2. Filled triangles are same ScI galaxies using single calibration. Again the distances are systematically too small with a single calibration. Error bars are rms uncertainty derived from errors in magnitudes and rotational velocities.